**Single electron transistor at room temperature fabricated on the large area surface of hydrogenated diamond like carbon thin film**

Nihar R. Ray[*,‖], Jagannath Datta[§], Hari S. Biswas[¶], Pintu Sen[†] and Kousik Bagani[‖]

[*]Center for Nano-Science and Surface Physics Project, [‖]Surface Physics Division, Saha Institute of Nuclear Physics, Kolkata -700 064, India

[§]Analytical Chemistry Division, BARC, Variable Energy Cyclotron Center, Kolkata-700 064, India

[¶]Department of Chemistry, Surendranath College, Kolkata-700 009

[†]Department of Physics, Variable Energy Cyclotron Center, Kolkata-700 064, India

**Introduction**

The single electron transistor (SET) switches on and off every time an electron is added to conducting island, called coulomb blockade (CB), isolated from two metallic leads, source (S) and drain (D), by insulator, called tunnel barrier (TB)[1]. In order to avoid thermally induced random tunneling events, the minimum size of CB has to be smaller than ~ 1 nm for room temperature (RT) operation of SET[2]. The present conventional microfabrication process limits the size of CB within a few tens of nm[3, 4]. The nature of the CB -metallic nanostructures, semiconducting, or quantum dot – is irrelevant for creation of SET, in first order approximation[2]. Here we show SET at RT, operated on macroscopic system-the two-dimensional (2D) large area (0.012 m × 0.012 m) surface of our hydrogenated diamond like carbon (HDLC) thin film[5-9]. Our results, free from the limitation of lowering the size of CB in sub-nanometer range with the existing unprofitable process[3, 4], are consistent with the theoretical results[1, 10, 11], and will be relevant for real application[2] at RT in a practical size device.

*Corresponding author. E-mail: niharranjan.ray@saha.ac.in

**Experiments**

The experimental data of SET at RT, based upon diamond like carbon (DLC) is not reported in the literature[12-14]. The HDLC thin film grown onto large area substrate Si (100), behaves as a 2D graphitic film, having both p-type and n-type surface conduction[9].

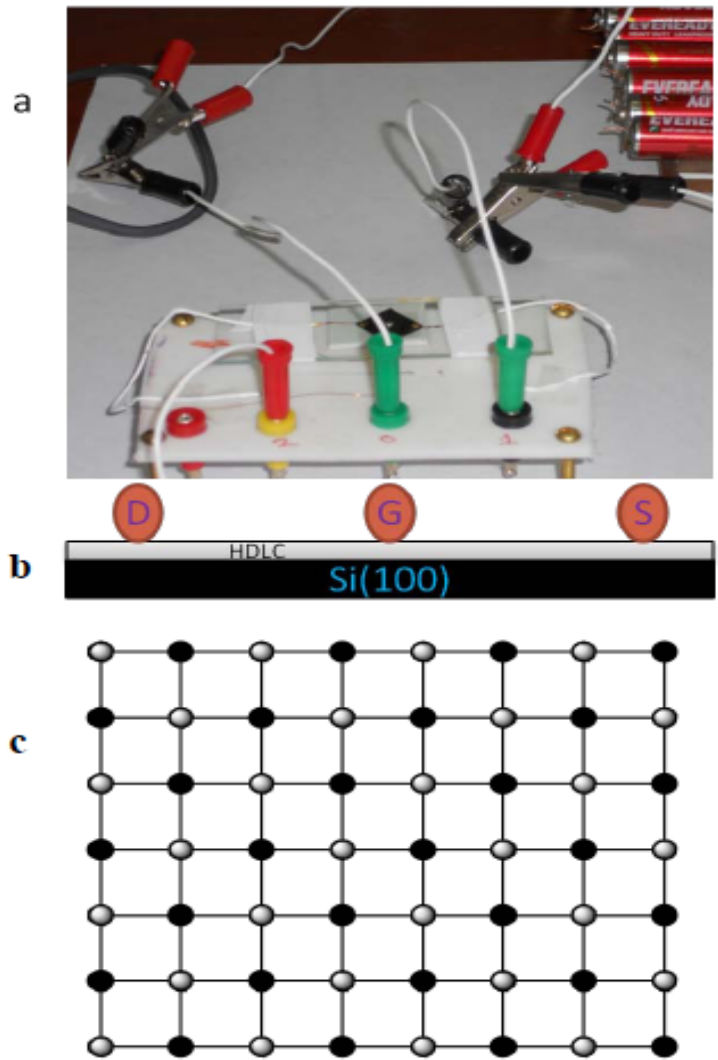

**Figure 1. Picture and Schematic layout of sample. a, A picture of three ohmic contacts on the surface of HDLC thin film and the connecting wires from these contacts. b, A schematic view of layers of Si(100) substrate, thin film of HDLC and three ohmic contacts: source (S), gate (G) and drain (D), corresponding to three**

**contacts in the picture, onto the surface of HDLC sample. c, A schematic view of the top of HDLC surface; solid circles represent non-conducting (sp$^3$ C-H) carbons, hollow circles represent conducting (sp$^2$ C=C) carbons, and the solid lines represents interconnections between solid and hollow circles indicating coherent domains of sp$^2$ and sp$^3$.**

A picture of ohmic contacts of Cu- source, gate and drain onto the HDLC surface[9, 14] (see Fig. 1a), layout of HDLC (see Fig. 1b), and schematic view of the top HDLC surface (see Fig. 1c) are shown. The top view shows coherent domains of sp$^3$ C-H (non-conductor) and sp$^2$ C=C (conductor) carbons and thus indicating resistive nature of the HDLC surface[15]. A Keithley 2635 source meter is used to measure the sheet resistance and I-V characteristics on the HDLC surface[9] at RT (295K). A confocal Micro-Raman spectrometer is used to measure Raman spectra of the HDLC samples[5-9]. In order to reduce the size of the domain of conducting (sp$^2$ C=C) carbons isolated by the domains of non-conducting (sp$^3$C-H) carbons, we have applied electrochemical method[16] for hydrogenation of sp$^2$ C=C bond on the pristine HDLC surface, and thus produce electrochemically hydrogenated diamond like carbon (ECHDLC) surface.

**Results and Discussion**

The typical sheet resistance of the ECHDLC surface ≈ 5-8 GΩ for sourcing current 1000 nA is much larger than that ≈ 75 MΩ of pristine HDLC surface for sourcing current 100 nA. The typical I-V characteristics between source and drain for pristine HDLC and ECHDLC surface (see Fig. 2a) respectively are consistent with the results of sheet resistances. The I-V characteristics (see Fig. 2a) show no hysteresis during measurements by increasing and decreasing drain voltage. The typical first-order Raman spectra of pristine HDLC and ECHDLC samples (see Fig. 2b) give $sp^3$ content, estimated from the position of G-peak ($\omega_G$, $\mu m^{-1}$) using the equation, $sp^3$ content = 0.24 – 48.9 ($\omega_G$ – 0.1580), derived in our earlier works[6]. Enhancement of $sp^3$ C-H content, from 47% (for pristine HDLC) to 49% (for ECHDLC) is observed. Moreover, we observe, increase of the slope of photoluminescence background[17] in the typical Raman spectrum of ECHDLC sample w.r.t that of pristine HDLC (see Fig. 2b) which clearly indicates increase of bonded hydrogen. Estimation of the ratio of slope (m) and intensity of G-peak ($I_G$) of the Raman spectra i.e. $m/I_G$ gives approximately 1.5% increase of $sp^3$ C-H bonds in the ECHDLC sample. The indigenous technique for enhancement of hydrogenation of pristine HDLC surface, and hence to reduce the size of the isolated domains of conducting carbons ($sp^2$ C=C), is reported in the present work for the first time.

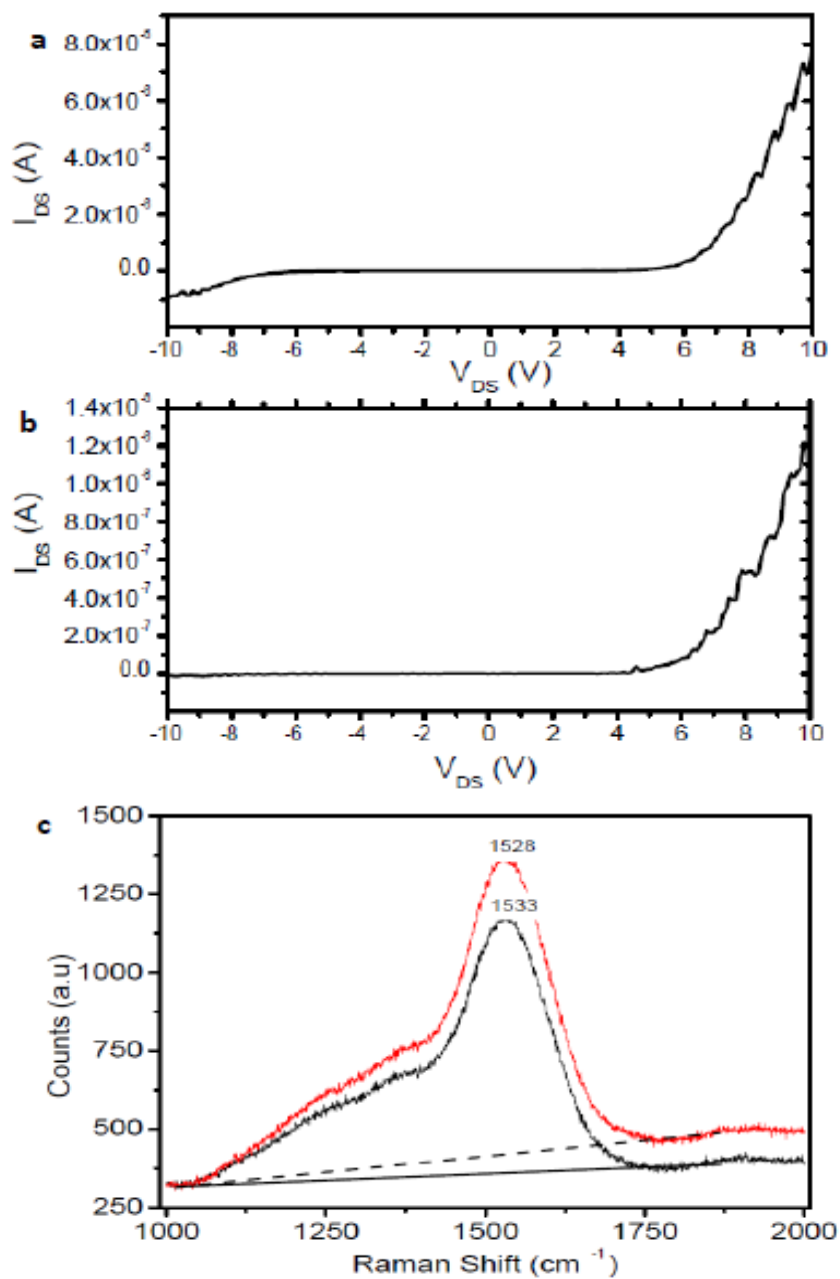

**Figure 2. Current (I) vs. Voltage (V) characteristic and Raman spectrum of sample. a,** Typical current ($I_{DS}$) vs. voltage ($V_{DS}$) characteristic of the pristine HDLC surface of area 0.012 m × 0.012 m. **b,** Typical current ($I_{DS}$) vs. voltage($V_{DS}$) characteristic of the ECHDLC surface of area 0.012 m × 0.012 m. **c,** Typical first order Raman spectra of pristine HDLC (black) and ECHDLC (red) samples.

The typical results on drain-source current ($I_{DS}$) vs. drain-source voltage ($V_{DS}$)/ time (t), and drain-source conductance ($\sigma_{DS}$) vs. drain-source voltage ($V_{DS}$) at different gate voltage ($V_G$) respectively, are shown in Figure 3. In the pristine HDLC surface, we observe quasi-periodic oscillation of hole current ($I_{DS}^h$) and conductance ($\sigma_{DS}^h$) for $V_G$ = +7.5 v (see Fig. 3a and b) and $V_G$ = +9 v (see Fig. 3c and d) respectively with "step-like" structure in I-V/conductance characteristic; the conductance value is approximately three orders lower than quantum conductance ($e^2/h$), and increases with $V_{DS}$. Similarly, in the ECHDLC surface, we observe quasi-periodic oscillation of electron current ($I_{DS}^e$) and conductance ($\sigma_{DS}^e$) for $V_G$ = -7.5 v (see Fig. 3e and f) and $V_G$ =-9 v (see Fig. 3g and h) respectively with "step-like" structure in I-V/conductance characteristic; here the conductance value is approximately two orders lower than $e^2/h$, and decreases with $V_{DS}$. For the estimation of single electron/hole capacitance ($C_e$, / $C_h$) and no. of charges (N) transferred to the CB, we measure increase of voltage ($\Delta V$) between two peaks, say peak i to peak (i+1) in $I_{DS}$ vs. $V_{DS}$ graphs, and current increase ($\Delta I$) with time ($\Delta t$) corresponding to peak (i+1) in $I_{DS}$ vs. t graphs in Fig. 3. From these measured values, we estimate, $\Delta Q = \Delta I \times \Delta t = Ne$; single electron/hole capacitance ($C_e / C_h$) = $e/\Delta V$.

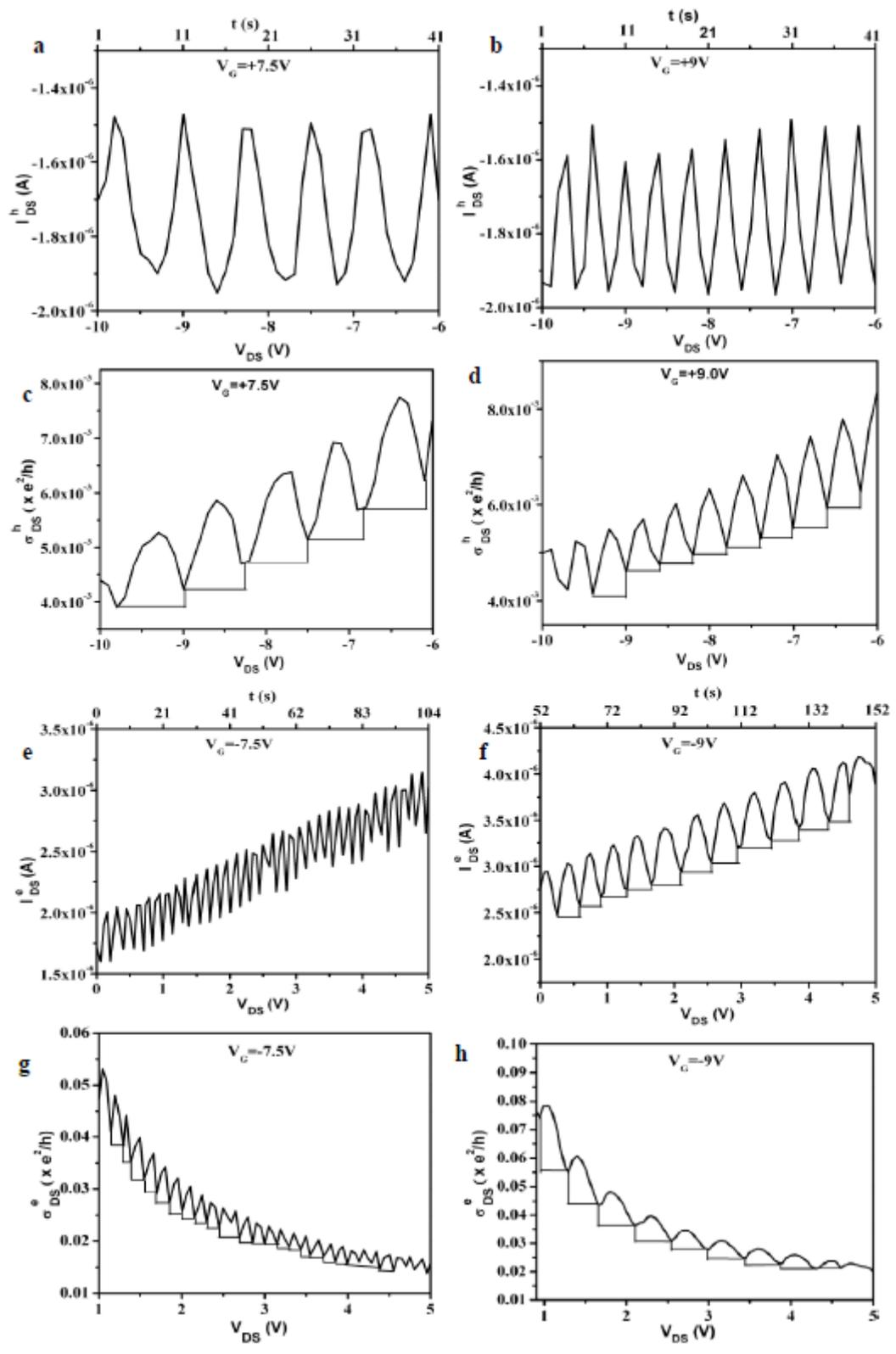

**Figure 3.** Current ($I_{DS}$) vs. Voltage ($V_{DS}$)/Time (t) and Conductance ($\sigma_{DS}$) vs. Voltage ($V_{DS}$) for a given Gate Voltage ($V_G$). a, $I_{DS}^h$ vs. $V_{DS}$ and $I_{DS}^h$ vs. t for $V_G$ = +7.5 v. b, $\sigma_{DS}^h$ vs. $V_{DS}$ for $V_G$ = +7.5 v. c, $I_{DS}^h$ vs. $V_{DS}$ and $I_{DS}^h$ vs. t for $V_G$ = +9.0 v. d, $\sigma_{DS}^h$ vs. $V_{DS}$ for $V_G$ = +9.0 v. e, $I_{DS}^e$ vs. $V_{DS}$ and $I_{DS}^e$ vs. t for $V_G$ = -7.5 v. f, $\sigma_{DS}^e$ vs. $V_{DS}$ for $V_G$ = -7.5 v. g, $I_{DS}^e$ vs. $V_{DS}$ and $I_{DS}^e$ vs. t for $V_G$ = -9.0 v. h, $\sigma_{DS}^e$ vs. $V_{DS}$ for $V_G$ = -9.0 v.

**Table 1 Electrical parameters for SET at RT**

| Parameter | Material | | | |
|---|---|---|---|---|
| | Pristine HDLC surface | | ECHDLC surface | |
| | $V_G$ = +7.5 V | $V_G$ = +9 V | $V_G$ = -7.5 V | $V_G$ = -9 V |
| $N_h$ | 0.825E+13 | 0.451E+13 | - | - |
| $N_e$ | - | - | 0.585E+13 | 1.056E+13 |
| $C_h$ (F) | 2E-19 | 4E-19 | - | - |
| $C_e$ (F) | - | - | 1E-18 | 6E-19 |
| $\Delta V$ (V) | 0.810 | 0.401 | 0.158 | 0.300 |
| $\Delta I$ (A) | 0.432E-6 | 0.355E-6 | 0.450E-6 | 0.552E-6 |
| $\Delta t$ (s) | 3.06 | 2.03 | 2.08 | 3.06 |

In Table 1 we give measured electrical parameters relevant for SET at RT on the surface of 2D HDLC thin film. In order to give a plausible physical explanation for the measured results in Table 1, we may assume a simple model: between S and D separated by macroscopic distance (~ 0.01 m) on 0.012 m × 0.012 m HDLC surface, there are large numbers of conducting ($sp^2$ C=C) carbon as CB separated by large numbers of non-conducting ($sp^3$ C-H) carbon as TB and CB and TB exist coherently[8] (see Fig. 1c). The increase of voltage $\Delta V$ between S and D for a given gate voltage $V_G$, when a single electron (hole) starts from S, passing through all CBs and TBs with finite tunneling probability ultimately reaches D, there is a peak in current/conductance. The next electron does not start from S until the tunneling of previous electron is completed, and hence there is a time-gap between the electrons reaching at D, resulting in a fall of current in I-V graph during that time-gap. Now, if we assume, total numbers of CBs / TBs are $10^{13}$, and each CB contributes only one conducting electron, then total equilibrium number of conducting electrons in all CBs, between S and D, will be $10^{13}$. Now during electron (hole) tunneling process between S and D with finite probability, the increase of number of electron in each CB will be $(1 \pm \varepsilon)$, and that in all CBs will be $(1 \pm \varepsilon) 10^{13}$. If we assume $0 < \varepsilon < 1$, then we can say that the measured values of $N_h$ and $N_e$ (see Table 1), are reasonable. Therefore our results, that periodic peaks in I-V and conductance characteristics correspond to average number of electrons in a single CB is not an integer but $(1 \pm \varepsilon)$, are consistent with theory[1,10,11] wherein average number of electrons in a single CB is (N+1/2). Moreover, height of the steps in I-V/conductance characteristics, depending upon gate voltage $V_G$, are not equal (see Fig. 3) but that for single CB, the corresponding steps are equal[18]. The values of $C_e$ are a few times larger than that of $C_h$ (see Table 1), but the order of magnitude for both values tells us that the diameter of CB is ~ 1 nm or less and the corresponding charging energy of several hundred meV (see Table 1) seems to be plausible[2]. Therefore the measured parameters ($N_h$, $N_e$, $C_h$, $C_e$, $\Delta V$ & $\Delta I$) in Table 1 are reasonable for SET at RT according to basic theory of SET[1, 2, 10, 11]. Considering the result $\Delta t$, that a single electron travels large numbers of CBs/TBs in time of seconds (see Table 1), we can say that sensitivity for detection of 'single electron event' or that for 'counting of electrons' should be high in our SET at RT. Thus our SET at RT seems to be relevant for practical applications in the future development of solid-state electrometer[19], electron pumps[20], artificial neural network[2, 21].

**Conclusion**

We conclude that charge quantization at RT and hence SET at RT on the large surface area of our 2D HDLC thin film, wherein there are ~ ten-trillion CBs separated by approximately same numbers of TBs with sub-nanometer size of each, and corresponding capacitance in the range of attofarad, respectively, is produced and operational.

……………………………………………………………………………………………

**Acknowledgements**

One of the authors (NRR) thanks Dept. of Atomic Energy, Govt. of India for funding during XI plan period to create the experimental facilities for carrying out the present work. We thank Mr. S. S. Sil and Mr. U. S. Sil for technical help and Dr. S Chattopadhyay and Dr. S Bhuia for fruitful scientific discussions.